\documentclass[aps,jcp,superscriptaddress,amsmath,amssymb,twocolumn]{revtex4-1}

\usepackage[utf8]{inputenc}
\usepackage{amsmath}
\usepackage{comment}
\usepackage{mathtools}
\usepackage{xcolor}
\usepackage[unicode]{hyperref}
\usepackage{revsymb}
\usepackage{tikz}
\usepackage{siunitx}
\usepackage{multirow}
\hypersetup{
   unicode=true,          % non-Latin characters in Acrobat??s bookmarks
   plainpages=false,
   colorlinks=true,       % false: boxed links; true: colored links
   linkcolor=blue,          % color of internal links
   citecolor=blue,        % color of links to bibliography
}
\usepackage{url}

\usepackage{dcolumn}
\newcolumntype{d}[1]{D{.}{.}{#1}}

\newcommand{\bos}[1]{\boldsymbol{#1}}
\def\Eh{E_\text{h}}
\def\cm{\text{cm}^{-1}}

\def\br{\boldsymbol{r}}

\def\iim{\text{i}}
\def\eem{\text{e}}
\def\tT{\text{T}}

\def\dd{\text{d}}
\def\epsi{\varepsilon}

\def\nel{n_\text{el}}

\def\XpSup{\text{X}\ ^2\Sigma_\text{u}^+}
\def\atSup{\text{a}\ ^3\Sigma_\text{u}^+}
\def\ctSgp{\text{c}\ ^3\Sigma_\text{g}^+}
\def\som{Supplementary Material}

\def\Eh{E_\text{h}}
\def\cm{\text{cm}^{-1}}
\def\br{\boldsymbol{r}}
\def\iim{\text{i}}
\def\Nb{N_\text{b}}

\def\nnuc{N_\text{nuc}}

\def\MV{\text{MV}}
\def\Done{\text{D1}}
\def\Dtwo{\text{D2}}
\def\OO{\text{OO}}

\def\bs{\boldsymbol{s}} %<-> spin vs. ECG shift...

\def\hbp{\hat{\boldsymbol{p}}}
\def\hbs{\hat{\boldsymbol{s}}} %<-> spin vs. ECG shift...
\def\br{\boldsymbol{r}}
\def\bnabla{\boldsymbol{\nabla}}

\def\bA{\boldsymbol{A}}
\def\ubA{\underline{\boldsymbol{A}}}
\def\hH{\hat{H}}

\def\vphi{\varphi}
\def\brho{\bos{\rho}}
\def\el{\text{el}}
\newcommand{\pd}[2]{\frac{\partial #1}{\partial #2}}
\def\dd{\text{d}}

\definecolor{ao}{rgb}{0.0, 0.5, 0.0}

\usepackage{setspace}
\usepackage[unicode]{hyperref}
\usepackage{soul}
\hypersetup{
   unicode=true,          % non-Latin characters in Acrobat??s bookmarks
   plainpages=false,
   colorlinks=true,       % false: boxed links; true: colored links
   linkcolor=black,          % color of internal links
   linkcolor=blue,          % color of internal links
   citecolor=blue,        % color of links to bibliography
   urlcolor=blue           % color of external links
}
\begin{document}

\title{Rovibrational computations for He$_2^+$ $\XpSup$\ including non-adiabatic, relativistic and QED corrections}
\author{Edit Mátyus}
\email{edit.matyus@ttk.elte.hu}
\affiliation{MTA–ELTE Lendület `Momentum' Molecular Quantum electro-Dynamics Research Group,
Institute of Chemistry, Eötvös Loránd University, Pázmány Péter sétány 1/A, Budapest, H-1117, Hungary}

\author{Ádám Margócsy}
\affiliation{MTA–ELTE Lendület `Momentum' Molecular Quantum electro-Dynamics Research Group,
Institute of Chemistry, Eötvös Loránd University, Pázmány Péter sétány 1/A, Budapest, H-1117, Hungary}

\date{\today}

\begin{abstract}
\noindent
We report the potential energy curve, the diagonal Born-Oppenheimer, non-adiabatic mass, relativistic, and leading-order quantum-electrodynamical (QED) corrections for the ground electronic state of the helium dimer cation; the higher-order QED and finite-nuclear size effects are also estimated.
The computations are carried out with an improved error control and over a broader configuration range compared to earlier work [D. Ferenc, V. I. Korobov, and E. Mátyus, Phys. Rev. Lett. 125, 213001 (2020)]. As a result, all rovibrational bound states are reported with an estimated accuracy of 0.005~$\cm$. \\[0.25cm]
\end{abstract}

\maketitle

%%%%%%%%%%%%%%%%%%%%%%%%%%%%%%%%%%%%%%%%%%%%%%%%%%%%%%%%%%%%%%%%%%%%%%%%%%%%%%%%%%%%%%%%%%%%%%%%%
%
% Introduction
%
%%%%%%%%%%%%%%%%%%%%%%%%%%%%%%%%%%%%%%%%%%%%%%%%%%%%%%%%%%%%%%%%%%%%%%%%%%%%%%%%%%%%%%%%%%%%%%%%%
\section{Introduction}
\noindent
Few-electron molecules and ions, such as He$_2^+$, have been at the focus of precision physics for many years. 
A highly accurate determination of energy intervals of small, `calculable' molecules can help test the small, formerly neglected physical effects and may lead to the refinement of physical constants.

Contrary to the ground state of $\text{He}_2$ (atoms held together only by van der Waals forces), the bond in the ground state of $\text{He}_2^+$ has a strong covalent character; as a consequence, this simple three-electron molecule has a rich rotational-vibrational spectrum, suitable for investigations in precision spectroscopy.

Experimental and theoretical studies of He$_2^+$ are almost as old as quantum mechanics itself; the first experimental determination of the equilibrium structure and dissociation energy of its ground electronic state goes back to the early 1930's \cite{Weizelbook1931}. Since then, experimental techniques underwent a great evolution, providing highly accurate results for the rovibrational intervals of He$_2^+$~$\XpSup$; reaching various highly vibrationally excited rotational states via microwave spectroscopy with fragmentation in an electric field \cite{CaPyKn95} or via ionization of He$_2$ Rydberg states, most notably, rovibrational states of $\atSup$~\cite{JaSeMe16,SeJaMe16,JaSeMe16vib,JaSeMe18,SeJaCaMeScMe20,SemeriaPhD2020,Jansen_habil2020}. The precision of these experiments is impressive for this small homonuclear diatomic molecule, the smallest uncertainties of rovibrational intervals are on the order of $\sim10^{-4} \, \text{cm}^{-1}$, or even less.
Since the first submission of this manuscript, comprehensive experimental data (of an overall 0.1~$\cm$ uncertainty) were published on the bound rovibrational intervals of He$_2^+$ \cite{HoWiShBeMe25} by using ionization pathways not only through the $\atSup$ but also the $\ctSgp$ electronic state of He$_2$ \cite{HoWiShBeMe25}.

Theoretical efforts towards the accurate description of the X$^+$ state have had a similarly long history, starting with the variational computations of Majorana in 1931~\cite{Ma31} and Pauling in 1933 \cite{Pa33}; results have constantly and significantly improved since then \cite{ReBrMa63,AcHo91,CaPyKn95,CeRy95,BaDa96,CeRy00,TuPaAd12,Ma18he2p,FeKoMa20,GePrGrTo23}.

The goal of this paper is to further improve the accuracy of the computed rovibrational intervals \cite{Ma18he2p,FeKoMa20}. The improvement is achieved via a more refined (more tightly optimised) non-relativistic Born-Oppenheimer energy and wave function. Using these, the computation of relativistic, quantum electrodynamics (QED), and non-adiabatic corrections is carried out with improved numerical error control. 
In the following sections, first, the theoretical framework is introduced (Sec.~\ref{sec:theory}), then the technical details of the computations are specified (Sec.~\ref{sec:comp}). Finally, all bound rovibrational energies are reported and compared with experimental data already available in the literature (Sec.~\ref{sec:rovib}) and earlier computations~\cite{TuPaAd12,Ma18he2p,FeKoMa20}.

%%%%%%%%%%%%%%%%%%%%%%%%%%%%%%%%%%%%%%%%%%%%%%%%%%%%%%%%%%%%%%%%%%%%%%%%%%%%%%%%%%%%%%%%%%%%%%%%%
%
% Theoretical framework
%
%%%%%%%%%%%%%%%%%%%%%%%%%%%%%%%%%%%%%%%%%%%%%%%%%%%%%%%%%%%%%%%%%%%%%%%%%%%%%%%%%%%%%%%%%%%%%%%%%
\section{Theoretical framework \label{sec:theory}}
\noindent %
The theoretical formalism is based on the Born-Oppenheimer (BO) non-relativistic quantum mechanics framework of electrons and atomic nuclei, \emph{e.g.,} \cite{PrMHBook84}. Both the Born-Oppenheimer (BO) and non-relativistic approximations are corrected by perturbation theory.
We start with the non-relativistic molecular Schrödinger equation
\begin{align}
  \hat{H} \Psi(\br,\brho) &= E \Psi(\br,\brho) \; ,
\end{align}
where the translationally invariant non-relativistic Hamiltonian for a homonuclear diatomic molecule is
\begin{align}
  \hat{H} =
    -\frac{1}{2\mu} \nabla^2_{\bos{\rho}}
    + \hat{H}_\el(\brho)
    +\frac{1}{8\mu} \hat{P}^2_\el \ 
\end{align}
with the total electronic momentum, $\hat{\boldsymbol{P}}_\el=-\iim\sum_{i=1}^{n_\el} \bnabla_{\br_i}$, 
the $\brho=\boldsymbol{R}_{1}-\boldsymbol{R}_{2}$ internuclear separation vector,  
the $\mu=M_\text{nuc}/2$ nuclear reduced mass, 
and the electronic Hamiltonian,
\begin{align}
  \hat{H}_{\text{el}}(\brho)
  =& 
  -\frac{1}{2} 
  \sum_{i=1}^{\nel} \bos{\Delta}_{\br_i} 
  -\sum_{i=1}^{\nel}\left[\frac{Z}{\left|\br_i-\frac{1}{2}\brho\right|}
  +\frac{Z}{\left|\br_i+\frac{1}{2}\brho\right|}\right] \nonumber \\
  &+ 
\sum_{i=1}^{\nel}\sum_{j=i+1}^{\nel}
  \frac{1}{\left|\br_i-\br_j\right|} 
  +\frac{Z^2}{\rho}
  \ .
  \label{Hel}
\end{align}

We start with solving the electronic Schr\"odinger equation, then compute corrections for the BO separation of the electronic and nuclear motions, the diagonal BO correction and the nuclear kinetic energy (mass) correction \cite{Te03,PaSpTe07,MaTe19}. Furthermore, we append the electronic energy with relativistic, QED, and finite nuclear size corrections. Relativistic nuclear effects are neglected in this work.
Finally, we solve the rovibrational Schrödinger equation using the corrected potential energy curve (PEC), $W$, and the effective, coordinate-dependent rotational and vibrational reduced masses, $\mu^\text{rot}(\rho)$ and $\mu^\text{vib}(\rho)$. This step amounts to solving the following radial equation for every $N$ rotational quantum number,
\begin{align}
    \Big[
    -
    \frac{\partial}{\partial \rho}
    \frac{1}{2 \mu^{\text{vib}}}
    \frac{\partial}{\partial \rho}
    +
    \frac{N(N+1)}{2 \mu^{\text{rot}}\rho^2}
    +
    W
    \Big]
    \psi_n(\rho)
    =
    E_n
    \psi_n(\rho)
    \; .
  \label{eq:radNSch}
\end{align}
In the following subsections, we define all corrections considered in this work and explain technical and computational details. 

%%%%%%%%%%%%%%%%%%%%%%%%%%%%%%%%%%%%%%%%%%%%%%%%%%%%%%%%%%%%%%%%%%%%%%%
% Electronic energy and wave function
%%%%%%%%%%%%%%%%%%%%%%%%%%%%%%%%%%%%%%%%%%%%%%%%%%%%%%%%%%%%%%%%%%%%%%%
\subsection{The non-relativistic electronic energy and wave function \label{sec:sch_ele}}
\noindent %
The electronic Schrödinger equation, 
\begin{align}
 \hat{H}_{\text{el}}(\brho)\vphi(\br,\brho)
 =
 U(\rho)\vphi(\br,\brho) \; ,
 \label{eq:Sch_electronic}
\end{align}
is solved variationally by approximating the electronic wave function as a linear combination of $N_\text{b}$ basis functions, 
\begin{align}
 \varphi(\br)
 =
 \sum_{\mu=1}^{N_\text{b}}c_{\mu} \, \phi_{\mu}(\br) \ .
 \label{eq:ECGexpansion}
\end{align}
A single basis function is written as, 
\begin{align}
  \phi_\mu(\br)
  =
  {\cal{P}}_{G} \, \mathcal{A}\Big[f(\br;\bA_\mu,\bs_\mu)\chi_{S,M_S}(\boldsymbol{\theta}_\mu)\Big] \ .
 \label{eq:basisfunction}
\end{align}
The spatial part is described by floating explicitly correlated Gaussian functions (fECGs), \emph{e.g.,}~\cite{SuVaBook98,MaRe12,MiBuHoSuAdCeSzKoBlVa13},
\begin{align}
  f(\br;\bA_\mu,\bs_\mu)
  =
  {\cal{N}}_\mu\exp\left[%
    -(\br-\bs_\mu)^\tT \ubA_\mu (\br-\bs_\mu)
  \right] \ ,
  \label{fECG}
\end{align}
where ${\cal{N}}_\mu$ is the normalization factor, ${\ubA}_\mu=\bA_\mu\otimes \bos{I}_3$, $\bA_\mu$ is a symmetric positive definite matrix, $\boldsymbol{I}_3\in\mathbb{R}^{3\times3}$ is the three-dimensional unit matrix, and the $\bs_\mu\in\mathbb{R}^{3\nel}$ `shift vector' is the centre of the fECG in the $3\nel$-dimensional configuration space,
\begin{equation}
 \bs_\mu=
 \begin{bmatrix}
  \bs_{\mu,1} \\
  \bs_{\mu,2} \\
  \vdots \\
  \bs_{\mu,\nel}
 \end{bmatrix}
 \ .
\end{equation}
Since the nuclei are fixed along the body-fixed $z$ axis, $\bos{\rho}=(0,0,\rho)$, we restrict the shift vector also to the $z$ axis to describe the $\Sigma$ symmetry of the electronic state. The ungerade symmetry of $\Sigma_\text{u}$ functions is implemented by explicit projection with $\mathcal{P}_\text{G}$ \emph{(vide infra)}.

$\chi_{S,M_S}(\boldsymbol{\theta}_\mu)$ is the three-electron spin function with the total $S=1/2$ and $M_S=1/2$ spin quantum numbers (or $M_S=-1/2$ after interchanging the $\alpha$ and $\beta$ labels):
\begin{align}
 |{\Theta_{\frac{1}{2},+\frac{1}{2}}}\rangle
 &=
 \frac{1}{\sqrt{2}}( |{\alpha\alpha\beta}\rangle - |{\alpha\beta\alpha}\rangle) \\
 |{\Theta'_{\frac{1}{2},+\frac{1}{2}}}\rangle
 &=
 \frac{1}{\sqrt{6}}( |{\alpha\alpha\beta}\rangle + |{\alpha\beta\alpha}\rangle-2|{\beta\alpha\alpha}\rangle) \; .
\end{align}
It was shown in Ref.~\citenum{CeRy93} that, in principle, a single spin function is sufficient from the relevant $(S,M_S)$ subspace; nevertheless, we retain both functions (for an increased the flexiblity of the Ansatz). So, we write the normalised $S,M_S$ adapted three-electron spin function as
\begin{align}
  \chi_{S,M_S}(\theta_\mu)
  =
  \sin(\theta_\mu) \Theta_{S,M_S}
  +
  \cos(\theta_\mu) \Theta'_{S,M_S}
  \label{eq:spinfun}
\end{align}
with the free $\theta_\mu\in[0,2\pi)$ parameter. The anti-symmetrizing operator in Eq.~\eqref{eq:basisfunction} acts on both the spatial and the spin degrees of freedom (and its action is calculated along the lines of Ref.~\citenum{MaRe12}). 

This procedure was generalised to four electrons in Ref.~\citenum{paper1}. 
In Ref.~\citenum{paper3}, the $\theta$ parameterisation (following Ref.~\citenum{SuVaBook98}) in Eq.~\eqref{eq:spinfun} was replaced by linear combination coefficients of the elementary many-electron spin functions represented as $2^N$ spinors (vectors) for the evaluation of the spin-dependent Breit-Pauli matrix elements.

In Eq.~\eqref{eq:basisfunction}, $\mathcal{A}$ is the anti-symmetrization operator for the electrons, ${\cal{P}}_{G}$ is the spatial projector to the selected irreducible representation of the point group (here: $\Sigma_\text{u}^+$ of $D_{\infty \text{h}}$). The spatial symmetry operations are implemented by mapping the $\bos{s}$ shift vectors of the fECGs~\cite{MaRe12,Ma19review}.

Substituting Eq.~(\ref{eq:ECGexpansion}) into Eq.~(\ref{eq:Sch_electronic}) yields a generalized matrix eigenvalue equation,
\begin{align}
  \boldsymbol{H}\boldsymbol{c}
  =
  U\boldsymbol{S}\boldsymbol{c} \ ,
 \label{eq:mateq}
\end{align}
with $H_{\mu\nu}=\langle\phi_\mu|\hat{H}_\text{el}|\phi_\nu\rangle$, $S_{\mu\nu}=\langle\phi_\mu|\phi_\nu\rangle$.
We solve the linear variational problem for $c_\mu$, and at the same time, minimise the $U$ electronic energy with respect to the $\bA_\mu,\bs_\mu$ non-linear parameters of the spatial functions; we also allow the $\bos{\theta}_\mu$ spin parameter to be relaxed during the energy minimisation. For a new basis function, the $\bA_\mu,\bs_\mu$ and $\bos{\theta}_\mu$ parameters are generated according to the stochastic variational method \cite{SuVaBook98,MaRe12}
and refined with the Powell approach \cite{Po04}. 
The basis functions are defined and refined one-by-one, and this allows us to use the fast eigenvalue update upon the rank-1 update of a known matrix-eigenvalue problem \cite{SuVaBook98}.
The $c_\mu$ linear parameters are found by direct solution of the linear eigenvalue equation, Eq.~\eqref{eq:mateq}. The main advantage of using fECGs is that the matrix elements of all operators relevant for the non-relativistic molecular problem can be calculated analytically. All computations were performed using our in-house developed, variational fECG-based computer program, QUANTEN. 

The electronic Schrödinger equation is variationally solved at several points over the quantum-dynamically relevant interval of the internuclear distance. At the computed points, the electronic energy (BO PEC value) is appended with the diagonal BO correction (DBOC) and finite nuclear-size corrections, and the leading- and (estimated) higher-order relativistic and QED corrections,
\begin{align}
  W=U+U_{\text{DBOC}}+U_{\text{f.nuc.}}+U_{\text{rel+QED}} \ .
\end{align}
The corrections are defined in the following subsections; specific computational details are collected in Sec.~\ref{sec:comp}.

\subsection{Post-BO corrections} 
The diagonal Born-Oppenheimer correction (DBOC) is computed in the centre-of-mass frame of the nuclei, and we aim to exploit the symmetry properties of the electronic state (for a numerically precise evaluation). 
Hence, the correction is written as a sum of an electronic, a vibrational and an electronic angular momentum part,
\begin{equation}
  U_{\text{DBOC}}=
    \frac{1}{8\mu} 
    \langle \hat{\boldsymbol{P}}^2_\el \rangle
    -\frac{1}{2\mu}
    \left\langle \frac{\partial^2}{\partial \rho^2} \right\rangle
    +
    \frac{1}{2\mu\rho^2}\langle \hat{L}^2_x + \hat{L}^2_y \rangle \ ,
  \label{eq:dboc}
\end{equation}
where each term is computed as an expectation value of the operators with the non-relativistic wave function. 
$\hat{L}^2_{x}$ and $\hat{L}^2_{y}$ are the total electronic angular momentum operators squared for the $x$ and $y$ directions.

The vibrational and rotational mass corrections~\cite{Ma18nonad,MaTe19}, which account for the effect of the electronic states beyond the X$^+$ state on the rovibrational dynamics, the effective reduced mass (depending on the $\rho$ internuclear distance) is written as  
\begin{align}
  \frac{1}{2\mu^\text{x}(\rho)}
  =
  \frac{1}{2\mu}
  \left[%
    1- \frac{\delta m^\text{x}(\rho)}{2\mu}
  \right] \ 
\end{align}
with $\text{x} = \text{vib and rot}$, 
\begin{align}
  \delta m^\text{vib}
  &=
  4\left\langle 
    \pd{\vphi}{\rho_z} \Big|
      (\hat{H}_\el-U)^{-1} P^\perp
    \Big| \pd{\vphi}{\rho_z}
  \right\rangle \ ,
  \label{eq:mvib} \\
  \delta m^{\text{rot}}
  &=
  4\left\langle 
    \pd{\vphi}{\rho_\beta} \Big|
      (\hat{H}_\el-U)^{-1} P^\perp
    \Big| \pd{\vphi}{\rho_\beta}
  \right\rangle \ .
  \label{eq:mrot}
\end{align}
$P^\perp = 1-|\vphi \rangle\langle\vphi|$ is the projector onto the orthogonal complement of the electronic Hilbert space to X$^+$ and $\beta=x,y$. Over the dynamically important range of the nuclear coordinate, no crossing with electronically excited states \cite{GePrGrTo23} of the $P^\perp$ subspaces ($^2\Sigma_\text{u}^+$ and $^2\Pi_\text{u}$ for the vibrational and rotational corrections, respectively) was observed. If there were crossing electronic states of the relevant symmetries, it would be necessary to explicitly (variationally) couple them in the quantum dynamics, and only the effect of the remaining (non-crossing) electronic states could be accounted for perturbatively \cite{MaTe19,MaFe22nad}.

\subsection{Relativistic \& QED corrections}
Relativistic and QED effects are accounted for in the electronic part of the problem, and they are computed as the sum of three main terms,
\begin{align}
  U_{\text{rel+QED}}
  =
  U_{\text{rel}}+U_{\text{lQED}}+U_{\text{hQED}} \ .
  \label{eq:Ecentroid}
\end{align}
The leading-order ($\alpha^2\Eh$) relativistic correction $U_{\text{rel}}$ is obtained as the expectation value of 
the (spin-independent terms of the) Breit-Pauli Hamiltonian with the non-relativistic electronic wave function \cite{BeSabook75,sucherPhD1958,JeAdBook22}
\begin{align}
  U_\text{rel}
  =
  \alpha^2\langle \hH_\MV + \hH_\Done +  
  \hH_\OO + \hH_\Dtwo + \hH_\text{SS,c} \rangle \ ,
  \label{eq:Hbp}
\end{align}
where $\alpha$ is the fine-structure constant.
The well-known terms are repeated here for completeness: the mass-velocity term,
\begin{align}
  \hH_\MV
  =
  -\frac{1}{8} 
  \sum_{i=1}^{\nel} (\hbp_{i}^{2})^2 \ ,\  %
\end{align}
for short: $\hH_\MV = -\frac{1}{8} \hat{p}^4$;
the one-electron Darwin term,
\begin{align}
  \hH_\Done
  =
  \frac{\pi}{2}\hat{\delta}_1 \ \ , \ \ 
  \hat{\delta}_1=\sum_{i=1}^{\nel}
  \sum_{A=1}^{\nnuc}
    Z_A\delta(\br_{iA}) \ ,
\end{align}
the orbit-orbit interaction,
\begin{align}
  \hH_{\OO}
  =
  -\frac{1}{2}
  \sum_{i=1}^{\nel}\sum_{j=i+1}^{\nel}
    \left[%
      \frac{1}{r_{ij}}\hbp_{i}\hbp_{j} + \frac{1}{r_{ij}^{3}}(\br_{ij}(\br_{ij}\hbp_{j})\hbp_{i})
    \right] \ ,
\end{align}
the two-electron Darwin term,
\begin{align}
  \hH_{\Dtwo}
  =
  -\pi\hat{\delta}_2 \ \ , \ \ 
  \hat{\delta}_2=
  \sum_{i=1}^{\nel}\sum_{j=i+1}^{\nel}
    \delta(\br_{ij}) \ ,
\end{align}
and the Fermi contact part of the spin-spin interaction ($\hat{\boldsymbol{s}}_i=I(1)\otimes ...\otimes \boldsymbol{\sigma}(i)/2\otimes...\otimes I(\nel)$ with the $\boldsymbol{\sigma}$ Pauli matrix), 
\begin{align}
  \hat{H}_\text{SS,c} 
  =  
  -\frac{8\pi}{3} 
  \sum_{i=1}^{\nel}\sum_{j=i+1}^{\nel}
    \hbs_i \hbs_j \delta(\br_{ij}) \ ,
  \label{eq:FermiSSc}
\end{align}
which can be simplified over the anti-symmetric subspace of the Hilbert space to (Appendix \ref{appendix:SSCrel})
\begin{align}
  \hat{H}_\text{SS,c} 
  =  
  2\pi \hat{\delta}_2 \;.
  \label{eq:FermiSScD}
\end{align}

The leading-order ($\alpha^3\Eh$) QED correction also arises as an expectation value with the electronic eigenstate~\cite{JeAdBook22,sucherPhD1958,Ar57}, 
\begin{align}
  U_{\text{lQED}}
  =&\alpha^3
  \Bigg\langle \frac{4}{3}\left[\frac{19}{30}-2\ln(\alpha)-\ln(k_0) \right]\hat{\delta}_1 \nonumber \\
   &+ \left[\frac{164}{15}+\frac{14}{3}\ln(\alpha) \right]\hat{\delta}_2
    -\frac{7}{6\pi}
  {\cal{P}}\left(\frac{1}{r^3}\right)  \Bigg\rangle \ ,
  \label{eq:E3}
\end{align}
where $\ln(k_0)$ is the (state-specific, non-relativistic) Bethe logarithm, 
\begin{align}
 \ln(k_0)=\frac{\left\langle\hat{\boldsymbol{P}}_\el(\hat{H}_\text{el}-U)\ln(2|\hat{H}_\text{el}-U|)\hat{\boldsymbol{P}}_\el\right\rangle}{2\pi\langle\hat{\delta}_1\rangle} \ ,
 \label{eq:BLterm}
\end{align}
and ${\cal{P}}(1/r^3)$ is the Araki-Sucher distribution (with the Euler-Mascheroni constant $\gamma$),
\begin{align}
 {\cal{P}}\left(\frac{1}{r^3}\right)
 =
 \lim_{\epsilon\rightarrow0^+}\sum_{i=1}^{\nel}\sum_{j=i+1}^{\nel}\Bigg[&\frac{\Theta(r_{ij}-\epsilon)}{r_{ij}^3} \nonumber \\
 +&4\pi (\gamma+\ln(\epsilon))\delta(\boldsymbol{r}_{ij})\Bigg] \ .  
 \label{eq:ASterm}
\end{align}

As advocated \emph{e.g.,} in Refs.~\cite{PuPa06,KoPiLaPrJePa11,PuKoPa15,PuKoCzPa16}, the higher-order ($\alpha^4\Eh$) QED correction to the centroid energy is approximated by
\begin{align}
  U_\text{hQED}
  =
  \pi\alpha^4\left[\frac{427}{96}-2\ln (2)\right]
  \sum_{A=1}^{\nnuc}Z^2_A\sum_{i=1}^{\nel}\langle\delta(\br_{iA})\rangle \ ,
\end{align}
which is a straightforward extension of the $m\alpha(Z\alpha)^5$ QED correction for hydrogen-like atoms~\cite{BaBeFe53,Eides2001} to several electrons and nuclei. This radiative one-electron part of the $\alpha^4\Eh$ correction is expected to be the most important contribution (Table~II of Ref.~\cite{Pa06} or Ref.~\cite{PuKoCzPa16} for the $1 \ ^1S$, $2 \ ^1S$ states of $\text{He}$ and the ground state of $\text{H}_2$, respectively).

Although a nuclear effect, it is listed after the hQED electronic correction for its smallness: the finite nuclear size correction to the centroid energy is given by (see \emph{e.g.} Sec.~XVII of Ref.~\cite{Eides2001}, or Sec.~5.3.2 of Ref.~\cite{JeAdBook22})
\begin{align}
  U_\text{f.nuc.}
  =
  \frac{2\pi}{3}\alpha^2\sum_{A=1}^{\nnuc}Z_A\frac{{\cal{R}}^2_A}{\lambdabar^2_\text{C}}\sum_{i=1}^{\nel}\langle\delta(\br_{iA})\rangle \ ,
  \label{eq:fnuc}
\end{align}
where $\lambdabar_\text{C}=\hbar/(mc)=\alpha a_0$ is the reduced Compton wavelength of the electron and ${\cal{R}}^2_A$ is the mean-squared radius of nucleus $A$ (here: $\alpha^{2+}$).

%%%%%%%%%%%%%%%%%%%%%%%%%%%%%%%%%%%%%%%%%%%%%%%%%%%%%%%%%%%%%%%%%%%%%%%%%%%%%%%%%%%%%%%%%%%%%
% 
% Computational details
%
%%%%%%%%%%%%%%%%%%%%%%%%%%%%%%%%%%%%%%%%%%%%%%%%%%%%%%%%%%%%%%%%%%%%%%%%%%%%%%%%%%%%%%%%%%%%%
%
\begin{figure*}
  \includegraphics[scale=1.]{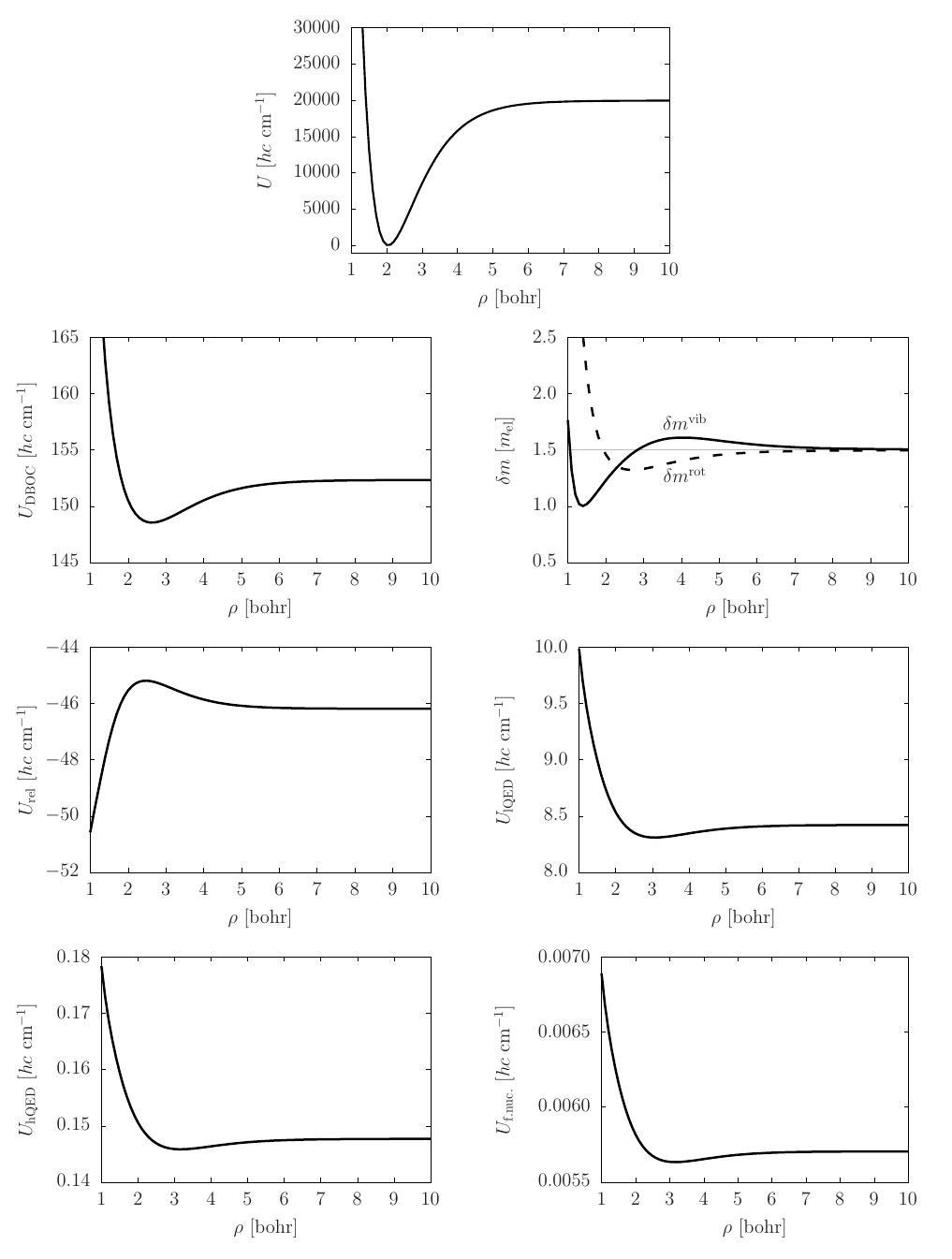}
\caption{%
  He$_2^+$~$\XpSup$: BO potential energy curve, $U$; diagonal BO correction, $U_\text{DBOC}$; rotational and vibrational mass corrections, $\delta m^\text{rot}$ and $\delta m^\text{vib}$; relativistic correction, $U_\text{rel}$; leading-order QED correction, $U_\text{lQED}$; and estimate for higher-order QED corrections, $U_\text{hQED}$, and the finite nuclear-size effect, $U_\text{f.nuc.}$. The PEC points (and the corrections) have been computed at several points in the $\rho\in[0.3,100]$~bohr interval; the computed dataset is available in the \som.
  \label{fig:PECcorr}
}
\end{figure*}
\section{Computational details \label{sec:comp}}
\paragraph{Converging the electronic energy; PEC generation}
The electronic energy was converged for the $\rho=2$~bohr internuclear distance (Table~\ref{tab:elconv}) by increasing the fECG basis set size and with extensive optimisation cycles of the non-linear basis function parameters. Since it is computed in a variational procedure, it is an upper bound to the exact electronic energy. Its value in the largest basis set ($\Nb=2250$) is estimated to be within (an upper bound) 10~n$\Eh$ of the exact electronic energy. 

\begin{table}%[h!]
  \caption{%
    He$_2^+$~$\XpSup$ ($\rho=2$~bohr):
    Convergence of the electronic energy, in $\Eh$,
    as a function of the $\Nb$ number of fECG basis functions, optimised in a variational procedure.
    The convergence of the total electronic orbital angular momentum squared, $\hat{L}_x^2$ and $\hat{L}_y^2$ expectation value is also shown in units of $\hbar^2$. 
    \label{tab:elconv}
  }
  \begin{tabular}{@{}rlll@{}}
    \hline\hline\\[-0.40cm]
      $\Nb$ & 
      \multicolumn{1}{c}{$U$} & &
      \multicolumn{1}{c}{$\langle\hat{L}_x^2\rangle=\langle\hat{L}_y^2\rangle$} \\
    \hline\\[-0.35cm]
    \multicolumn{3}{l}{$\rho=2.000$~bohr:} & \\    
           & --4.994 441 57   & \cite{TuPaAd12} & \\
      20   & --4.99           &              & 3.221 \\
      100  & --4.994          &              & 3.216 3 \\
      250  & --4.994 4        &              & 3.216 23 \\
      500  & --4.994 44       &              & 3.216 204 \\
      750  & --4.994 441      &              & 3.216 203 7 \\
      1000 & --4.994 441 6    &              & 3.216 203 3 \\
      1200 & --4.994 441 7    &              & 3.216 203 6 \\
      1500 & --4.994 441 74   &              & 3.216 203 63 \\
      1750 & --4.994 441 75   &              & 3.216 203 73 \\
      2000 & --4.994 441 76   &              & 3.216 203 76 \\
      2250 & --4.994 441 765  &  $^\text{a}$ & 3.216 203 82 \\
    \hline\\[-0.35cm]      
    \multicolumn{3}{l}{$\rho=2.042$~bohr:} & \\
           & --4.994 643 94   & \cite{TuPaAd12} & \\
           & --4.994 644 104  & \cite{FeKoMa20} & \\
      2250 & --4.994 644 141  & $^\text{b}$ & \\
    \hline\hline\\[-0.40cm]      
  \end{tabular}
  \begin{flushleft}
    $^\text{a}$~This basis parameter set was used for the PEC generation. \\
    $^\text{b}$~Rescaled from the 2250 basis set, followed by 10 full Powell refinement cycles.
  \end{flushleft}
\end{table}

\begin{table*}%[h!]
  \caption{%
    He$_2^+$~$\XpSup$ ($\rho=2$~bohr):
    Convergence of the expectation value of singular operators, $\hat{H}_\text{MV}$, $\hat{H}_\text{D1}$, $\hat{H}_\text{D2}$ in the relativistic correction and the ${\cal{P}}(1/r^3)$ term in the QED correction, with the electronic wave function. $\Nb$ is the number of the basis functions.
    Direct evaluation: no regularization; the integral transformation (IT) technique \cite{PaCeKo05,JeIrFeMa22} and the numerical Drachmanization approach \cite{RaFeMaMa24} were used. 
    The orbit-orbit term, $\langle \hat{H}_\text{OO}\rangle$, and the centroid relativistic energy, $U_\text{rel}$ in $\alpha^2\Eh$ units, are also shown.
    \label{tab:relHe2pX}
  }
    \centering
    \begin{tabular}{@{}r lll lll@{}}
    \hline\hline\\[-0.35cm]
       \multicolumn{1}{c}{$\Nb$} & 
       \multicolumn{1}{c}{$\langle \hat{H}_{\rm MV} \rangle$} &
       \multicolumn{1}{c}{$\langle \hat{H}_{\rm D1} \rangle$} &
       \multicolumn{1}{c}{$\langle \hat{H}_{\rm D2} \rangle$} &        
       \multicolumn{1}{c}{$\langle \hat{H}_\OO \rangle$} &   
       \multicolumn{1}{c}{$U_\text{rel}$} &
       \multicolumn{1}{c}{$\langle{\cal{P}}(1/r^3)\rangle$} \\
       \hline\\[-0.35cm]
       \multicolumn{7}{l}{Direct evaluation} \\  
100       &		$-$23.64     &	19.46    &	$-$0.388      &	$-$0.083 1    &	$-$3.872    & $-$ \\ 
250       &		$-$23.849    &	19.66    &	$-$0.380      &	$-$0.082 75   &	$-$3.888    & $-$ \\
500       &		$-$23.894    &	19.71    &	$-$0.377 6    &	$-$0.082 70   &	$-$3.891 0  & $-$ \\
750       &		$-$23.897    &	19.711   &	$-$0.376 3    &	$-$0.082 686  &	$-$3.892 6  & $-$ \\
1000      &		$-$23.913 7  &	19.727   &	$-$0.376 1    &	$-$0.082 683  &	$-$3.893 04 & $-$ \\
1200      &		$-$23.914 0  &	19.728 0 &	$-$0.376 0    &	$-$0.082 682  &	$-$3.892 80 & $-$ \\
1500      &		$-$23.913 8  &	19.728 0 &	$-$0.375 8    &	$-$0.082 680 4 &	$-$3.892 74 & $-$ \\
1750      &		$-$23.913 8  &	19.728 0 &	$-$0.375 78   &	$-$0.082 680 2 &	$-$3.892 74 & $-$ \\
2000      &		$-$23.913 81 &	19.728 0 &	$-$0.375 76   &	$-$0.082 680 1 &	$-$3.892 74 & $-$ \\
2250      &		$-$23.913 84 &	19.728 0 &	$-$0.375 75  &	$-$0.082 680 0 &	$-$3.892 76 & $-$ \\
\hline\\[-0.35cm]
       \multicolumn{7}{l}{IT regularization} \\
     20        & {$-$23.13}       &	{18.73}    	& {$-$0.440}   	& $-$ &	{$-$4.045}      & {1.00}  \\
     100       & {$-$23.925}      &	{19.729}   	& {$-$0.384}   	& $-$ &	{$-$3.895}      & {1.34}  \\
     250       & {$-$23.943}      &	{19.754}   	& {$-$0.377}   	& $-$ &	{$-$3.895 2}    & {1.44}  \\
     500       & {$-$23.943 3}    &	{19.755 3}	& {$-$0.375} 	    & $-$ &	{$-$3.895 53}   & {1.46}  \\
     750       & {$-$23.943 4}    &	{19.755 4}	& {$-$0.374 7}	& $-$ &	{$-$3.895 97}   & {1.469}  \\
     1000      & {$-$23.943 4}    &	{19.755 52}	& {$-$0.374 6}	& $-$ & {$-$3.896 02}   & {1.471}  \\
     1200      & {$-$23.943 40}   &	{19.755 53}	& {$-$0.374 5}	& $-$ &	{$-$3.896 00}   & {1.472}  \\
     1500      & {$-$23.943 40}   &	{19.755 54}	& {$-$0.374 53}	& $-$ &	{$-$3.896 01}   & {1.471 9} \\
     1750      & {$-$23.943 41}   &	{19.755 541}	& {$-$0.374 53}	& $-$ &	{$-$3.896 02}   & {1.471 7} \\
     2000      & {$-$23.943 416}  &	{19.755 543}	& {$-$0.374 524}	& $-$ &	{$-$3.896 030}  & {1.471 82} \\
     2250      & {$-$23.943 419}  &	{19.755 544}	& {$-$0.374 522}	& $-$ &	{$-$3.896 033}  & {1.471 89} \\
\hline\\[-0.35cm]
       \multicolumn{7}{l}{Numerical Drachman regularization} \\       
       20   &  $-$23.74      & 19.608      &	$-$0.363       &  $-$ & $-$3.85       & {1.54} \\
       100  &  $-$23.940     &	19.750      &	$-$0.373 9     &  $-$ & $-$3.899      & {1.48} \\
       250  &  $-$23.945     &	19.754 9    &	$-$0.374 39    &  $-$ & $-$3.898      & {1.475} \\
       500  &  $-$23.944 8   &	19.755 4    &	$-$0.374 46    &  $-$ & $-$3.897 7    & {1.474} \\
       750  &  $-$23.944 3   &	19.755 4    &	$-$0.374 480   &  $-$ & $-$3.897 0    & {1.473 8} \\       
       1000 &  $-$23.944 25  &	19.755 51   &	$-$0.374 482   &  $-$ & $-$3.896 94   & {1.473 7} \\
       1200 &  $-$23.944 22  &	19.755 511  &	$-$0.374 483   &  $-$ & $-$3.896 91   & {1.473 7} \\       
       1500 &  $-$23.944 13  &	19.755 5130 &	$-$0.374 484 8 &  $-$ & $-$3.896 815  & {1.473 691} \\
       1750 &  $-$23.944 120 &	19.755 5130 &	$-$0.374 485 1 &  $-$ & $-$3.896 802  & {1.473 687} \\
       2000 &  $-$23.944 114 & 19.755 5133 &	$-$0.374 485 3 &  $-$ & $-$3.896 796  & {1.473 685} \\
       2250 &  $-$23.944 106 & 19.755 5136 &   $-$0.374 485 4 &  $-$ & $-$3.896 787  & {1.473 683} \\
    \hline\hline
    \end{tabular}
\end{table*}

Starting from the basis parametrisation, which has been extensively optimised for $\rho=2$~bohr, we generated the potential energy curve by making consecutive small displacements of the nuclear geometry. After every geometry displacement, we have rescaled the $\bos{s}$ shift vectors of the fECGs in proportion to the geometry change and have carried out repeated refinement (Powell) cycles to optimise the basis parameters by minimisation of the energy for that nuclear geometry \cite{CeRy95,FeKoMa20,FeMa22h3}. We first computed the PEC points over the [0.3,10]~bohr interval with a 0.05~bohr step size. This set was extended with points up to 16.5~bohr with a 0.1~bohr step size, and then, (with an exploratory purpose) to 100.50~bohr with a 1~bohr step size.
As a result of the consecutive basis generation and reoptimization procedure, we think that the relative precision of the PEC in a `small neighbourhood', \emph{e.g.,} near the bottom of the PEC minimum, is smaller than the absolute convergence error.
At every PEC point, we computed post-BO corrections, relativistic and QED corrections, and also the finite-nuclear size correction. 

\vspace{0.5cm}
\paragraph{Post-BO corrections}
The DBOC was computed according to Eq.~\eqref{eq:dboc}. Table~\ref{tab:elconv} shows the convergence of the total electronic angular momentum terms; they converge relatively fast. The largest numerical uncertainty is due to the $\rho$ second derivative term, which is computed by numerical differentiation with respect to $\rho$. It is performed by rescaling the $\bos{s}$ shift vectors of the fECGs in proportion to the $\rho\pm\delta\rho$ variation, and it is sufficiently accurate (causing 0.000\ 1~$\cm$ or less numerical uncertainty in the rovibrational intervals). 

The non-adiabatic rotational and vibrational mass correction terms \cite{Ma18he2p,FeKoMa20} were converged by extensive variational optimisation of the auxiliary basis set representing the reduced resolvent in Eqs.~\eqref{eq:mvib} and \eqref{eq:mrot}. The auxiliary basis sets were of $\Sigma_\text{u}^+$ and $\Pi_\text{u}$ symmetry for the vibrational and rotational mass correction, respectively. The mass correction terms are estimated to be accurate to $5\cdot10^{-4}$~$m_\text{el}$ over the $\rho\in[0.3,10]$~bohr interval.

\vspace{0.5cm}
\paragraph{Relativistic and QED corrections}
The relativistic corrections, computed as the expectation value of the Breit-Pauli Hamiltonian, include several terms that are known for their (extremely) slow convergence in Gaussian basis sets towards the complete basis limit upon direct evaluation of the integrals \cite{Dr81,PaCeKo05,JeIrFeMa22,RaFeMaMa24}; among the leading QED corrections, the convergence of the Araki-Sucher term is similarly hindered for the same reason. 
Two alternative `regularization' schemes, by rewriting the original expressions with less singular operators, are used here to accelerate the convergence. The two regularization schemes are tested against one another.

In the integral transformation (IT) approach, the expectation value is rewritten in the form of an integral, and the integration is split into low-range and high-range regions by a separation parameter $\Lambda$; the low-range part is evaluated with numerical quadrature, while a semi-analytical asymptotic expansion is fitted to the high-range part (the relevant formulae of Ref. \cite{PaCeKo05} can be derived along the lines of  Ref.~\cite{JeIrFeMa22}; we note the factor of $2$ misprint in Eq.~(19) of Ref.~\cite{PaCeKo05}, corrected in Ref.~\cite{CePrKoMeJeSz12}). 

The so-called `Drachmanization' approach is based on the rewriting of expectation values with singular operators via identities which hold for the exact non-relativistic wave function \cite{Dr81,PaCeKo05}. The result is an expectation value with an operator that is much less localised around electron-nucleus or electron-electron coalescence points, leading to significantly improved basis set convergence. The Drachmanization approach was formerly feasible for ECGs centred at the origin (zero shift vectors), but molecular applications, which require fECGs, were hindered by incalculable integrals. We have recently proposed a numerical Drachmanization scheme (numDr) for $\langle\hat{H}_\text{MV}\rangle$, $\langle\hat{H}_\text{D1}\rangle$ and $\langle\hat{H}_\text{D2}\rangle$~\cite{RaFeMaMa24}, in which the missing integrals are evaluated by numerical integration. The numerical Drachmanization of $\langle{\cal{P}}(1/r^3)\rangle$ can be achieved along the same lines; details are given in the Appendix \ref{appendix:ASnumDr}.

According to our experience, the IT technique \cite{PaCeKo05,JeIrFeMa22} requires manual adjustment of the cutoff and fitting parameters, whereas the recently proposed numDr approach is more robust. Therefore, the convergence test was carried out using both techniques at $\rho=2$~bohr, but the PEC corrections were computed at all PEC points with the numDr technique. At $\rho=2$~bohr, we estimate the relativistic correction (numDr value) to be accurate to 0.005~$\cm$, and its relative error in the rotational and rovibrational intervals is thought to be even smaller. The error of the relativistic correction mainly stems from $\langle\hat{H}_\text{MV}\rangle$ which $-$ even in its Drachmanized form $-$ contains a term sensitive to the electron-electron cusp condition (see, \emph{e.g.} Ref.~\citenum{PuKoPa17}). This issue is in principle better handled in the IT approach (all cusp conditions being explicitly built into its asymptotic formulae), however, due to the aforementioned instabilities of the fitting procedure, we found numDr to be a safer choice for computing $\langle\hat{H}_\text{MV}\rangle$ along the PEC.

At the largest computed internuclear distance (100.5~bohr), all the above Drachmanized expectation values were also compared to the known exact values at dissociation limit, and a satisfactory agreement was found ($\langle\hat{H}_\text{MV}\rangle$ showing the largest, $\sim2\cdot10^{-4}$ discrepancy due to the above mentioned cusp problem, all other quantities accurate to $\sim3\cdot10^{-5}$ at least). In particular, the numerical value of the Araki-Sucher term was checked against the asymptotic formula
$
 \left\langle{\cal{P}}(r^{-3})\right\rangle=\left\langle{\cal{P}}(r^{-3})\right\rangle_{\text{He}}+2 \, \rho^{-3}+{\cal{O}}(\rho^{-4}) 
$
for $\rho>10 \, \text{bohr}$, where $\left\langle{\cal{P}}(r^{-3})\right\rangle_{\text{He}}\approx0.98927245$ is the exact atomic value \cite{Dr88}. The other three expectation values approach their dissociation limits as $\sim\text{const}\cdot\rho^{-4}$.

The Bethe logarithm is notorious for its computational difficulties; its direct evaluation would be inaccurate due to the importance of the highly excited electronic state contributions in Eq.~\eqref{eq:BLterm}. The approach of Schwartz~\cite{Sc61,KoHiKa13,Ko19,FeMa22bethe} circumvents this problem by reexpressing $\ln(k_0)$ as a principal value integral over the photon momentum. In this form, an auxiliary basis set can be optimised upon minimisation of a target functional (not the energy), which allows for systematic improvement of the Bethe logarithm itself. The auxiliary basis sets must be optimised for several values of photon momenta at each point along the PEC, which makes the Bethe logarithm evaluation computationally demanding. According to numerical experience, the Bethe logarithm value is only weakly dependent on ${\nel}$ and on the electronic state, and thus, it is often a good approximation to use the ion-core values of Ref.~\cite{FeKoMa20}. Hence, the Bethe logarithm of $\text{He}_2^+(^2\Sigma_\text{u}^+)$ is approximated with that of the single electron $\text{He}_2^{3+}(^2\Sigma_\text{g}^+)$. We note that this approximation was numerically checked (and confirmed) by explicit computation for the three-electron He$_2^+$ (X$^+$) at $\rho=2$~bohr in Ref.~\cite{FeMa22bethe}.

Figure~\ref{fig:PECcorr} shows the PEC and all correction curves (for the [1,10]~bohr interval; numerical values over [0.3,100.50]~bohr are available in the \som), which were used to define the radial equation, Eq.~\eqref{eq:radNSch}. For each $N$ rotational quantum number, the radial equation, Eq.~\eqref{eq:radNSch}, was solved using the discrete variable representation (DVR) \cite{LiCa00} with the $L^{(a)}_n$ ($a=2$) Laguerre polynomials scaled to the $[0.3,100]$~bohr interval. $N=500$ DVR points were used in the final computations; they were sufficient to converge all bound states to $10^{-5}$~$\cm$.
Finally, we note that the $^4$He nuclei ($\alpha^{2+}$) are spin-0 bosons, and hence, for the $\Sigma_\text{u}^+$ electronic state, the odd $N$ values satisfy the Pauli principle for the nuclei. Furthermore, since the helium-4 nuclei are spinless, nuclear spin-rotation and nuclear spin-spin couplings are absent. The computation of the coupling between the magnetic moments of the electron spin and molecular rotation (electron-spin-rotation coupling) is left for future work (as part of working towards the `relativistic recoil' corrections of this system).

The physical constants and conversion factors used in this work are 
according to CODATA~2022~\cite{codata22}, 
$\alpha = 7.2973525643(11)\cdot10^{-3}$, %
$M(^4\text{He}) = 7.29429954171(17)\cdot10^{+3}\ m_\text{el}$, %
$r(^4\text{He}) = 4.3467(5)\cdot10^{-3}\ \lambdabar_\text{C}$, %
$\Eh=2.1947463136314(24)\cdot10^{+5}\ (hc \, \text{cm}^{-1})$, 
$\Eh=6.579683920 4999(72)\cdot10^{+9}\ (h \, \text{MHz})$ (used to compare with the values of Ref.~\citenum{CaPyKn95}).

\begin{table*}%[h!]
  \caption{%
    He$_2^+$ $X^+, v=0$: rotational intervals, $\tilde\nu=\tilde\nu(0,N)-\tilde\nu(0,1)$ in cm$^{-1}$. The estimated uncertainty of the intervals computed in this work and listed in this table is 0.000\ 5~$\cm$.
    \label{tab:rotint}
  } 
  \scalebox{0.92}{%
  \begin{tabular}{@{}l@{}%
    d{4.6}@{\ }d{4.4}@{\ }d{4.4}@{\ } %
    d{5.4}@{\ }d{5.4}@{\ }d{6.4}@{\ } %
    d{6.4}@{\ }d{6.4}@{\ }d{6.5}@{}}
  \hline\hline\\[-0.35cm]
  \multicolumn{1}{r}{$N$:} &
  \multicolumn{1}{c}{3} &	
  \multicolumn{1}{c}{5} &
  \multicolumn{1}{c}{7} &
  \multicolumn{1}{c}{9} &
  \multicolumn{1}{c}{11} &
  \multicolumn{1}{c}{13} &
  \multicolumn{1}{c}{15} &
  \multicolumn{1}{c}{17} &
  \multicolumn{1}{c}{19} \\
  \hline\\[-0.35cm]
   $\tilde\nu$
     &  70.937\ 70 & 198.363\ 9 & 381.831\ 7 & 620.699\ 6 & 914.135\ 7 & 1\ 261.122\ 7 & 1\ 660.464\ 6 & 2\ 110.793\ 2 & 2\ 610.576\ 4	\\
   $\tilde\nu_\text{Expt.}^{\ \ast}-\tilde\nu$
     &  -0.000\ 11 &   0.000\ 8 &   0.002\ 9  &   0.002\ 5 &	 0.001\ 0 &	     0.001\ 5 &     -0.001\ 9 &	    0.000\ 0  &     -0.002\ 0 \\
   $\delta_\text{Expt.}^{\ \ast}$
     &  0.000\ 080 &   0.000\ 8 &   0.000\ 8  &   0.000\ 9 &	 0.000\ 8 &	     0.000\ 8 &      0.000\ 9 &	    0.000\ 9  &      0.000\ 9 \\
  \hline\\[-0.35cm]
  $\tilde\nu_\text{Expt.}^{\ \ast}-\tilde\nu_\text{Comp.\cite{FeKoMa20}}$ 
              &  -0.000\ 10	& 0.000\ 9 & 0.003\ 0 &	0.002\ 7 & 0.001\ 3 & 0.001\ 9 &  -0.001\ 3 & 0.000\ 7 &  -0.001\ 1 \\
  \hline\hline
  \end{tabular}
  }
  \begin{flushleft}
    \vspace{-0.15cm}
    \scalebox{0.95}{%    
    $^\ast$~The $\tilde\nu_\text{Expt.}$ experimental intervals with the $\delta_\text{Expt.}$ uncertainties are taken from Refs.~\citenum{SeJaMe16} and \citenum{SeJaCaMeScMe20}.}
  \end{flushleft}
  \caption{%
    He$_2^+$ $X^+, v=1$: %
    vibrational and rovibrational intervals, $\tilde\nu$ in cm$^{-1}$. The estimated uncertainty of the intervals computed in this work and listed in this table is 0.000\ 5~$\cm$.
    \label{tab:rvint}
  } 
  \scalebox{0.92}{%
  \begin{tabular}{@{}l d{6.6} |@{}c@{} d{5.5}d{6.5} d{6.4}d{6.4}d{6.5}@{}}
  \hline\hline\\[-0.35cm]
  & 
  \multicolumn{1}{c}{} &
  \multicolumn{5}{c}{$\tilde\nu(1,N)-\tilde\nu(0,1)$} \\
  \cline{4-8} \\[-0.55cm]
  \multicolumn{1}{r}{} &
  \multicolumn{1}{c}{\raisebox{0.25cm}{$\tilde\nu(1,0)-\tilde\nu(0,0)$}} & 
  \multicolumn{1}{r}{$N$:} &
  \multicolumn{1}{c}{1} &	  
  \multicolumn{1}{c}{3} &	
  \multicolumn{1}{c}{7} &
  \multicolumn{1}{c}{11} &
  \multicolumn{1}{c}{13} \\
  \hline\\[-0.35cm]
  $\tilde\nu$	
    &	 1628.383\ 4 && 1\ 627.931\ 6 & 1\ 696.610\ 3 & 1\ 997.562\ 9 & 2\ 512.689\ 3 &	2\ 848.370\ 4 \\
  $\tilde\nu_\text{Expt.}^{\ \ast}-\tilde\nu$ 
    &	    -0.000\ 2 &&      0.000\ 2 &     -0.000\ 7 &		 0.000\ 4 &	    -0.002\ 2 &	    -0.001\ 4 \\
  $\delta_\text{Expt.}^{\ \ast}$
    &	    0.001\ 2 &&      0.001\ 2 &      0.001\ 2 &		 0.001\ 2 &		 0.001\ 2 &      0.001\ 2 \\
  \hline\\[-0.35cm]  
  $\tilde\nu_\text{Expt.}^{\ \ast}-\tilde\nu_{\text{Comp.\cite{FeKoMa20}}}$
             &	 0.002\ 3	  &&     0.002\ 7  & 0.001\ 9	    &	   0.002\ 9 &	   0.000\ 4 &	   0.001\ 2 \\
  \hline\hline
  \end{tabular}
  }
  \begin{flushleft}
    \vspace{-0.15cm}
    \scalebox{0.95}{%
    $^\ast$~The $\tilde\nu_\text{Expt.}$ experimental intervals with the $\delta_\text{Expt.}$ uncertainties are taken from Refs.~\citenum{SeJaMe16} and \citenum{SeJaCaMeScMe20}.
    }
  \end{flushleft}
\end{table*}

\section{Rovibrational energies and comparison with experiment \label{sec:rovib}}
\noindent %
The new potential energy and correction curves improve upon previous computations in terms of the convergence of the (variational) electronic energy (BO PEC) and the numerical error control of all corrections. Regarding the corrections,  the same physical effects are accounted for (beyond the BO, non-relativistic framework) as in Ref.~\citenum{FeKoMa20}, but all details of the computations are carried out with a more mature methodology over a broader interval of the internuclear separation.
As a result, the rovibrational energies reported in this work (the full energy list is deposited in the \som) are expected to improve upon the results reported earlier \cite{TuPaAd12,Ma18he2p,FeKoMa20}. 
 
In this section, we compare the new results with experimental values already available in the literature. We use the $(v,N)$ notation to refer to a state with the $v$ vibrational label and $N$ rotational quantum number. 

The rotational intervals (Table~\ref{tab:rotint}) are in excellent agreement with experimental data reported in Refs.~\citenum{SeJaMe16} and \citenum{SeJaCaMeScMe20}; no significant numerical improvement is seen over the computed (lowest-energy) rotational intervals of Ref.~\citenum{FeKoMa20}. 

The vibrational fundamental and rovibrational intervals (Table~\ref{tab:rvint}) improve upon Ref.~\citenum{FeKoMa20}, and are now in excellent agreement with the experimental values of Ref.~\citenum{SeJaMe16} (and Ref.~\citenum{HoWiShBeMe25}). The improvement is the result of the more accurate PEC representation (primarily), but also the more precise corrections play a role; a better numerical error control is achieved for all corrections compared to Ref.~\citenum{FeKoMa20}. 

There is experimental data \cite{CaPyKn95} also for the high vibrationally ($v=22,23$) and rotationally ($N=1,3$) excited states close to the dissociation limit. Table~\ref{tab:highex} shows earlier computed and experimental intervals. No or little improvement is seen here; the formerly available computations \cite{TuPaAd12,Ma18he2p} were already in good agreement with experiment. In the earlier computations \cite{TuPaAd12,Ma18he2p}, the (missing) small physical corrections and the convergence error of the PEC seem to cancel for these close-lying states.
For the tight convergence of the PEC and the careful evaluation of the physical corrections reported in this work, we think that the remaining deviation (`This Work' in the table) from experiment is dominantly due to the (electron) spin-rotation effect, which is not included in the computations, but present in the experimental data in Table~\ref{tab:highex}. The missing spin-rotation effect is estimated to be 0.002~$\cm$, which is indicated as the uncertainty of the computed values.
We note that in Tables~\ref{tab:rotint} and \ref{tab:rvint}, the empirical data (Expt.), derived from the experiment, are without the spin-rotation contributions.

\begin{table}
  \caption{%
    Highly excited bound-state energy intervals near the dissociation limit, $\tilde\nu$ in cm$^{-1}$.
    \label{tab:highex}
  }
  \begin{tabular}{@{} l d{7.7} @{\ \ \ \ } d{7.7} @{}}
    \hline\hline\\[-0.35cm]
    & 
    \multicolumn{1}{c}{$\tilde\nu(23,3)-\tilde\nu(22,5)$} & 
    \multicolumn{1}{c}{$\tilde\nu(23,3)-\tilde\nu(23,1)$} \\
    \hline\\[-0.35cm]
    Comp.~\cite{TuPaAd12}  & 5.260	     & 2.002	\\
    Comp.~\cite{Ma18he2p}  & 5.250	     & 1.999	\\
    This work       & 5.2488(20)   & 2.0008(20) \\
    Expt. \cite{CaPyKn95}  & 5.248277(7) & 2.00055(7) \\
    \hline\hline
  \end{tabular}
\end{table}

\begin{figure}
  \includegraphics[width=8cm]{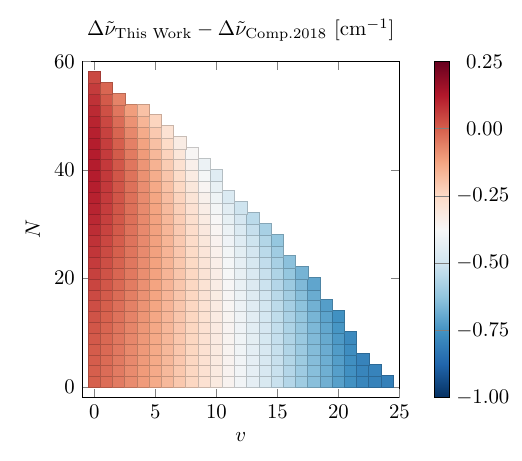}
  \caption{%
    All bound states of He$_2^+$ $\XpSup$ computed in this work improve upon earlier results reported in Ref.~\citenum{Ma18he2p} (Comp.2018) referenced to the zero-point vibrational energy (ZPVE) in both computations, 
    $\Delta \tilde\nu(v,N)=\tilde\nu(v,N)-\tilde\nu(0,0)$.
    The ZPVE has improved by
    $\tilde\nu(0,0)_\text{This Work}-\tilde\nu(0,0)_\text{Comp.2018}=0.799\ 8~\cm$.
    \label{fig:compare}
  }
\end{figure}

Finally, Fig.~\ref{fig:compare} compares all bound rovibrational states with those computed in Ref.~\citenum{Ma18he2p} with non-adiabatic mass corrections and using the PEC and DBOC correction of Ref.~\citenum{TuPaAd12}, but without relativistic and QED corrections, and a lower level of convergence and error control. For several higher rotationally or vibrationally excited states, a significant improvement, up to 1.00~$\cm$, is observed in this work.
This improvement was necessary to observe an excellent
overall agreement with the comprehensive experimental dataset (of 0.1~cm$^{-1}$ uncertainty) that has been published \cite{HoWiShBeMe25} since the first submission of this manuscript.

%%%%%%%%%%%%%%%%%%%%%%%%%%%%%%%%%%%%%%%%%%%%%%%%%%%%%%%%%%%%%%%%%%%%%%%%%%%%%%%%%%%%%%%%%%%%%%
%
%
%
%%%%%%%%%%%%%%%%%%%%%%%%%%%%%%%%%%%%%%%%%%%%%%%%%%%%%%%%%%%%%%%%%%%%%%%%%%%%%%%%%%%%%%%%%%%%%%
%\clearpage
\section{Summary, conclusion, outlook \label{sec:conclusion}}
\noindent%
This paper has reported the theoretical description of the ground electronic state of He$_2^+$ accounting for small, non-adiabatic, relativistic, and QED effects as in Ref.~\cite{FeKoMa20}, but with improved error control, and over a broader internuclear separation interval, $\rho\in[0.3,100]$ bohr. All in all, the current rovibrational energy list is expected to be the most comprehensive and most accurate to date.

The current rovibrational intervals are expected to be accurate to  0.005~$\cm$. 
The currently largest source of uncertainty is due to the numerical error control of the considered terms, most importantly, the convergence of the electronic energy and wave function (which can be further improved, if necessary), the convergence of the singular relativistic corrections, and the ion-core approximation of the Bethe logarithm. We think that the largest neglected physical effect is the non-adiabatic relativistic coupling, which can be computed in the future by considering the coupling terms of the two contact transformations corresponding to the non-adiabatic and relativistic parts of the problem. Furthermore, we plan to compute the magnetic coupling due to the magnetic moments of the electron-spin and molecular rotation, which has already been measured experimentally to high precision~\cite{JaSeMe18}.

\vspace{0.25cm}
\section{Acknowledgement}
\noindent Financial support of the European Research Council through a Starting Grant (No.~851421) is gratefully acknowledged. We also thank the Momentum (Lendület) Program of the Hungarian Academy of Sciences (LP2024-15/2024). 
We thank DKF for awarding us access to the Komondor HPC facility based in Hungary.

%\bibliography{references}

\vspace{0.25cm}
\section*{Appendix} 

\subsection{On the operator relation between the two-electron Fermi contact and delta function terms \label{appendix:SSCrel}}
\noindent
Eq. (\ref{eq:FermiSScD}) was proven in Exercise~2.2 of Ref.~\cite{HeJoOlbook} in a second-quantized formalism based on spin-orbitals.
For the explicitly correlated, first quantized formalism of the present paper, 
one may want to further elaborate on this relation. The wave function has the general structure
\begin{align}
 \Psi(\boldsymbol{r})&=\frac{1}{\sqrt{{\nel}!}}\sum_{{\cal{P}}\in S_{{\nel}}}(-1)^{\pi_{\cal{P}}}\langle\boldsymbol{r}|{\cal{P}}|\Phi\rangle \, {\cal{P}}_\text{spin} \,\chi \nonumber\\
 &=\frac{1}{\sqrt{{{\nel}}!}}\sum_{{\cal{P}}\in S_{{\nel}}}(-1)^{\pi_{\cal{P}}}\langle{\cal{P}}^{-1}\boldsymbol{r}|\Phi\rangle \, {\cal{P}}_\text{spin} \, \chi \ ,
\end{align}
$\boldsymbol{r}$ standing for $\boldsymbol{r}_1,...,\boldsymbol{r}_{\nel}$, and $\chi$ being some combination of spinor product functions; usually, $\chi$ is chosen to be a linear combination of degenerate spin eigenfuncions (as discussed in Sec. \ref{sec:sch_ele}), although this is not necessary. The action of the operator $\delta(\br_{ij})$ sets $\boldsymbol{r}_i=\boldsymbol{r}_j$, so due to antisymmetry, the only nonvanishing terms in $\chi$ must be of the form $ \alpha_i\otimes\beta_j-\beta_i\otimes\alpha_j$ (the other ${{\nel}}-2$ spinor factors unaffected by $\delta(\boldsymbol{r}_{ij})$ and their coefficients are not shown).
\newline
When acting on $\Psi$ with not only $\delta(\br_{ij})$ but $\delta(\br_{ij}) \, \hat{\boldsymbol{s}}_i\cdot\hat{\boldsymbol{s}}_j$, the $\hat{\boldsymbol{s}}_i\cdot\hat{\boldsymbol{s}}_j$ part acts on this antisymmetric spinor product. Using
\begin{equation}
 \hat{\boldsymbol{s}}_i\cdot\hat{\boldsymbol{s}}_j=\hat{s}_{iz}\hat{s}_{jz}+\frac{1}{2}\Big(\hat{s}_{i+}\hat{s}_{j-}+\hat{s}_{i-}\hat{s}_{j+}\Big) \ ,
\end{equation}
we immediately find
\begin{align}
 \hat{\boldsymbol{s}}_i\cdot\hat{\boldsymbol{s}}_j(\alpha_i\otimes\beta_j-\beta_i\otimes\alpha_j)=-\frac{3}{4}(\alpha_i\otimes\beta_j-\beta_i\otimes\alpha_j) \ ,
\end{align}
meaning
\begin{equation}
  \delta(\boldsymbol{r}_{ij}) \, \hat{\boldsymbol{s}}_i\cdot\hat{\boldsymbol{s}}_j \, \Psi(\boldsymbol{r})
 =-\frac{3}{4} \delta(\boldsymbol{r}_{ij}) \, \Psi(\boldsymbol{r}) \ ,
\end{equation}
and
\begin{equation}
 \hat{H}_\text{SS,c}\, \Psi(\boldsymbol{r})=2\pi \, \hat{\delta}_2\, \Psi(\boldsymbol{r}) \ .
\end{equation}
Apart from antisymmetry under simultaneous interchanges in coordinate and spin space, nothing was assumed of $\Psi$ or more specifically, $\chi$; in general, $\chi$ does not need to be an eigenfunction of either $\hat{\boldsymbol{S}}^2$, or $\hat{S}_z$. For this reason, $\hat{H}_\text{SS,c}=2\pi \, \hat{\delta}_2$ can be stated as an operator equality on the fully antisymmetric subspace of the $\nel$-particle Hilbert space.

\iffalse
First, we note that the spatial part of the operator is $\delta(\br_{12})$. Furthermore, we write the many-electron spin function as a linear combination of elementary spin functions. Then, we consider the matrix representation of the permutation ($\hat{P}_{12}$) and spin ($\hat{S}_{12}=\hbs_1\hbs_2$) operators over the $\alpha\alpha,\alpha\beta,\beta\alpha,\beta\beta$ basis set, 
\begin{align}
  P^\text{spin}_{12} 
  =
  \left(%
    \begin{array}{rrrr}
      1 & 0 & 0 & 0 \\
      0 & 0 & 1 & 0 \\
      0 & 1 & 0 & 0 \\
      0 & 0 & 0 & 1 \\
    \end{array}
  \right)
\end{align}
and 
\begin{align}
  S_{12}
  =
  \frac{1}{4}
  \left(%
    \begin{array}{rrrr}
      1 &  0 &  0 & 0 \\
      0 & -1 &  2 & 0 \\
      0 &  2 & -1 & 0 \\
      0 &  0 &  0 & 1 \\
    \end{array}
  \right)  \; ,
\end{align}
then,
\begin{align}
  S_{12}-S_{12}P^\text{spin}_{12} 
  =
  -\frac{3}{4}(I-P^{\text{spin}}_{12}) \; ,
\end{align}
which can be used to prove the relation, Eq.~\eqref{eq:FermiSScD}.
\fi

\subsection{Numerical Drachmanization of the Araki-Sucher term \label{appendix:ASnumDr}}
\noindent%
Following Ref. \cite{PaCeKo05}, the Drachmanized representation of the Araki-Sucher term is
\begin{align}
  \left\langle{\cal{P}}\left(\frac{1}{r^3}\right)\right\rangle
  &=
  4\pi(1+\gamma)\left\langle\hat{\delta}_2\right\rangle+
  2\left\langle {\hat{\cal{L}}}_{\text{ee}}(U-\hat{V})\right\rangle \nonumber \\
  &-\sum_{k=1}^{\nel}\left\langle\overset{\leftarrow}{\nabla}_k {\hat{\cal{L}}}_{\text{ee}}\overset{\rightarrow} {\nabla}_k\right\rangle \ ,
  \label{eq:ASDr}
\end{align}
where $U$ is the BO energy, $\hat{V}=\hat{V}_\text{ee}+\hat{V}_\text{eN}+\hat{V}_\text{NN}$,
\begin{align}
  \hat{{\cal{L}}}_{\text{ee}}
  =
  \sum_{i=1}^{\nel}\sum_{j=i+1}^{\nel}\frac{\text{ln}(r_{ij})}{r_{ij}} \ ,   
  \label{eq:Lee}
\end{align}
and $\overset{\leftarrow}{\nabla}$ refers to differentiation acting to the left. This representation gives rise to integrals in the $\hat{\mathcal{L}}_{\text{ee}}(U-\hat{V})$ terms which cannot be calculated in a closed analytical form for general floating ECG basis functions. As advocated in Ref. \cite{RaFeMaMa24}, a robust and efficient numerical integration scheme can be found via the integral identity 
\begin{align}
  r^{-\alpha}
  =
  \frac{2}{%
    \Gamma(\frac{\alpha}{2})
  }
  \int_{-\infty}^{+\infty}
    \dd s\ 
    \eem^{-r^2 \eem^{2s} + \alpha s} \; ,
  \label{eq:sint}
\end{align}
where $\Gamma(z)$ is Euler's Gamma function; Eq.~\eqref{eq:sint} can be verified with the substitution $s=\frac{1}{2}\ln(t/r^2)$.
Setting $\alpha=1+\epsi$ and expanding both sides around $\varepsilon=0$ to first order with
\begin{align}
  \Gamma\left(\frac{1+\epsi}{2}\right)
  =
  \sqrt{\pi}
  -
  \sqrt{\pi}\left(\ln (2)+\frac{\gamma}{2}\right)\varepsilon+{\cal{O}}(\varepsilon^2)
\end{align}
yield
\begin{align}
  \frac{1}{r} 
  = 
  \frac{2}{\sqrt{\pi}}
  \int_{-\infty}^{+\infty}
    \dd s\ 
    \eem^{-r^2 \eem^{2s} + s} \ ,
  \label{eq:oner}
\end{align}
and
\begin{align}
  \frac{\ln (r)}{r}
  =
  -\frac{2}{\sqrt{\pi}}
    \int_{-\infty}^{+\infty}
      \dd s\ 
      \eem^{-r^2 \eem^{2s} + s}
      \left(%
      s + \ln (2)+\frac{\gamma}{2}
      \right) \ .
  \label{eq:lnroner}
\end{align}
After substituting the latter expression in $\langle\hat{\cal{L}}_{\text{ee}}(\hat{V}_\text{ee}+\hat{V}_\text{eN})\rangle$, the $r$-integrations can be carried out analytically in the fECG representation, while the (rapidly converging) $s$-integration is done numerically.
The required fECG matrix elements are of the form
\begin{align}
 V_{\mu\nu,kl,ij}(s)
 &=
 \left\langle f_\mu\left|\frac{1}{r_{kl}} %
   \eem^{-\eem^{2s} r_{ij}^2}
   \right|f_\nu\right\rangle \ ,
\end{align}
and (for $A=1,...,\nnuc$)
\begin{align}
  V_{\mu\nu,kA,ij}(s)
  &=
  \left\langle f_\mu\left|\frac{1}{r_{kA}} %
  \eem^{-\eem^{2s} r_{ij}^2}
  \right|f_\nu\right\rangle \ ,
\end{align}
for which the detailed expressions were already discussed in \emph{e.g.,} the Supplementary Material of Ref.~\citenum{RaFeMaMa24}. 
\newline
The fECG matrix elements of $\hat{{\cal{L}}}_\text{ee}$ and $\overset{\leftarrow}{\nabla}\hat{{\cal{L}}}_\text{ee}\overset{\rightarrow}{\nabla}$ can be calculated fully analytically, without any need for numerical integration. For $\hat{{\cal{L}}}_\text{ee}$, we have
\begin{widetext}
\begin{align}
 \langle f_\mu|\hat{{\cal{L}}}_\text{ee}|f_\nu\rangle
 =
 S'_{\mu\nu}\sum_{i=1}^{\nel}\sum_{j=i+1}^{\nel}\frac{1}{t_{\mu\nu,ij}^{1/2}}\Bigg[\frac{\ln\left({t_{\mu\nu,ij}}\right)-\gamma}{2}{\cal{F}}_0(x_{\mu\nu,ij})+{\cal{G}}_0(x_{\mu\nu,ij})\Bigg] \ ,
\end{align}
\end{widetext}
where $S'_{\mu\nu}=\langle f_\mu|f_\nu\rangle$, 
\begin{equation}
 t_{\mu\nu,ij}=(\boldsymbol{e}_i-\boldsymbol{e}_j)^T(\bA_\mu+\bA_\nu)^{-1}(\boldsymbol{e}_i-\boldsymbol{e}_j) \ ,
\end{equation}
and
\begin{equation}
 x_{\mu\nu,ij}=\sqrt{\frac{(\vec{s}_{\mu\nu,i}-\vec{s}_{\mu\nu,j})^2}{t_{\mu\nu,ij}}} \ ,
\end{equation}
with the $\boldsymbol{e}_i$ unit vectors on $\mathbb{R}^{\nel}$, and $\vec{s}_{\mu\nu,i}$ is the 3-vector associated with the $k$-th electron in
 $\bs_{\mu\nu}\in\mathbb{R}^{3\nel}$:
\begin{equation}
 \bs_{\mu\nu}=(\underline{\bA}_\mu+\underline{\bA}_\nu)^{-1}\Big[\underline{\bA}_\mu\bs_\mu+\underline{\bA}_\nu\bs_\nu\Big] \ ,
\end{equation}
with $\underline{\boldsymbol{X}}=\boldsymbol{X}\otimes\boldsymbol{I}_3\in\mathbb{R}^{3\nel\times3\nel}$.
The two special functions are
\begin{equation}
 {\cal{F}}_0(x)=\frac{\text{erf}(x)}{x} \ ,
\end{equation}
and
\begin{equation}
 {\cal{G}}_0(x)=-\frac{1}{\sqrt{\pi}}\frac{\partial}{\partial a} {_1F_1}\left(a,\frac{3}{2},-x^2\right)\Bigg|_{a=\frac{1}{2}} \ ,
 \label{eq:G0}
\end{equation}
where ${_1F_1}(a,b,z)$ is the confluent hypergeometric function. Since ${\cal{G}}_0(x)$ has a convergent power series for all $x$, the first $\sim100$ Taylor coefficients can be used to set up a compact Pad\'e approximant representation, accurate to $\sim10^{-15}$ for $x<6$. For larger $x$, an asymptotic expansion can be used (exponentially small terms neglected):
\begin{equation}
 {\cal{G}}_0(x)\sim\frac{1}{x}\left[\ln(x)+\frac{\gamma}{2}-\frac{1}{2}\sum_{p=1}^K\frac{(2p-1)!!}{2^{p}p}x^{-2p}\right] \ ,
\end{equation}
with an error at most $\sim10^{-15}$ for $x>6$ (using $K=20$).
\newline
For matrix elements of $\overset{\leftarrow}{\nabla}\hat{{\cal{L}}}_\text{ee}\overset{\rightarrow}{\nabla}$, we have
\begin{widetext}
\begin{align}
 \sum_{k=1}^{\nel}\langle\nabla_k f_\mu|\hat{{\cal{L}}}_\text{ee}|\nabla_k f_\nu\rangle=S'_{\mu\nu}\sum_{\beta=0}^2\sum_{i=1}^{\nel}\sum_{j=i+1}^{\nel}C^{(\beta)}_{\mu\nu,ij}\frac{1}{t_{\mu\nu,ij}^{\beta+1/2}}\left[\frac{\ln(t_{\mu\nu,ij})-\gamma}{2}{\cal{F}}_\beta(x_{\mu\nu,ij})+{\cal{G}}_\beta(x_{\mu\nu,ij})\right] \ ,
\end{align}
where
\begin{align}
C_{\mu\nu,ij}^{(0)}&=3\text{Tr}[\widehat{\bA}_{\mu\nu}]-\boldsymbol{w}_{\mu\nu}^T\boldsymbol{w}_{\mu\nu} \ , \\
 C_{\mu\nu,ij}^{(1)}&=-6(\boldsymbol{e}_i-\boldsymbol{e}_j)^T(\bA_\mu+\bA_\nu)^{-1}\bA_\mu \bA_\nu (\bA_\mu+\bA_\nu)^{-1}(\boldsymbol{e}_i-\boldsymbol{e}_j)-2\boldsymbol{w}_{\mu\nu}^T\boldsymbol{\chi}_{\mu\nu} \ ,\\
 C_{\mu\nu,ij}^{(2)}&=2(\vec{s}_{\mu\nu,i}-\vec{s}_{\mu\nu,j})^2-\boldsymbol{\chi}_{\mu\nu,ij}^T\boldsymbol{\chi}_{\mu\nu,ij} \ ,
\end{align}
with
\begin{equation}
 \widehat{\bA}_{\mu\nu}=2\bA_\nu(\bA_\mu+\bA_\nu)^{-1}\bA_\mu \ \ , \ \ 
 \boldsymbol{w}_{\mu\nu}=\underline{\widehat{\bA}}_{\mu\nu}(\bs_\mu-\bs_\nu) \ \ , \ \  
 \boldsymbol{\chi}_{\mu\nu,ij}=(\underline{{\bA}}_{\mu}-\underline{{\bA}}_{\nu})(\underline{{\bA}}_{\mu}+\underline{{\bA}}_{\nu})^{-1}\underline{{\boldsymbol{J}}}_{ij}\bs_{\mu\nu} \ ,
\end{equation}
$\boldsymbol{J}_{ij}=(\boldsymbol{e}_i-\boldsymbol{e}_j)(\boldsymbol{e}_i-\boldsymbol{e}_j)^T$ being rank 1, and
\begin{align}
 {\cal{F}}_1(x)=\frac{\sqrt{\pi} \, \text{erf}(x)-2x\exp(-x^2)}{2\sqrt{\pi} \, x^3} \ \ \ &, \ \ \ {\cal{G}}_1(x)=-\frac{{\cal{G}}_0'(x)}{2x} \ , \\
 {\cal{F}}_2(x)=\frac{3\sqrt{\pi} \, \text{erf}(x)-2x(2x^2+3)\exp(-x^2)}{4\sqrt{\pi} \, x^5} \ \ \ &, \ \ \ 
 {\cal{G}}_2(x)=\frac{x{\cal{G}}_0''(x)-{\cal{G}}_0'(x)}{4x^3} \ .
\end{align}
\end{widetext}
We note that all of these functions have finite $x\rightarrow0$ limits. Accurate short- and long-range representations of ${\cal{G}}_1$ and ${\cal{G}}_2$ can be constructed analogously to the case of ${\cal{G}}_0$, Eq.~\eqref{eq:G0}.

\end{document}